# Common errors in Generative AI systems used for knowledge extraction in the climate action domain.

Denis Havlik
Cooperative Digital Technologies
Austrian Institute of Technology (AIT)
Vienna, Austria

Marcelo Pias[1]
Computer Science Centre
Federal University of Rio Grande - FURG
Rio Grande, Brazil

**Abstract:** *Large Language Models (LLMs) and, more specifically, the Generative Pre-Trained Transformers (GPT) can help stakeholders in climate action to explore digital knowledge bases, extract and utilize climate action knowledge in sustainable manner. However, LLMs are "probabilistic models of knowledge bases" that excel at generating convincing texts but cannot be entirely relied upon due to the probabilistic nature of the information produced. This brief report illustrates the problem space with examples of LLM responses to some of the questions of relevance to the climate action.*

## 1. Introduction

Designed initially as engines that put one word after another based on statistical probability, Large Language Models (LLMs) and, more specifically, the Generative Pre-Trained Transformers (GPT) (Devlin, 2018) (Ashish Vaswani, 2017) (OpenAI, 2023) are now capable of generating, classifying, and summarizing documents due to their emerging abilities (Wei, 2022). If utilized wisely, the LLMs can be instrumental in tackling sustainability challenges where users explore digital knowledge bases to undertake document analyses for climate action (M. Pathak, 2022).

It is important to note that the current Language Model Models (LLMs) are not equivalent to expert knowledge databases due to limitations in language construction, as discussed in (Fabio Petroni, 2019). According to Thomas G. Dietterich[2], "LLMs are probabilistic models of knowledge bases." Although the text generated by these models may appear convincing, it cannot be entirely relied upon due to the probabilistic nature of the information produced.

This brief report illustrates some of the critical risks and issues that computing researchers should address before the climate change research community can safely depend on LLMs to extract knowledge from climate-related documents.

## 2. The problem space

This section discusses various types of errors and risks commonly associated with generative AI. It highlights the potential societal dangers that could arise due to the uncritical use of generative AI outputs in the context of climate change. It is worth noting that all the tests outlined in this section were conducted between September and November 2023.

### Risk 1: Incomplete answers

Most knowledge extraction work focuses on tasks that involve qualitative information, such as answering questions about the main topics covered in a document. However, extracting factual data from climate-related text remains an unexplored and challenging area (Webersinke, 2022) (Thomas Diggelmann, 2020). Objective data refers to evidence and facts rather than personal interpretations or theories. Extracting more empirical knowledge, driven by specific indicators and target values (factual data), is crucial for providing helpful tools to users.

We conducted a simple test to explore the potential of Language Models for extracting factual data. We prompted three chat systems with the question in **prompt-1** (Table 1).

---

[1] Work developed while co-author visited the Austrian Institute of Technology from July to November 2023.

[2] "What's wrong with LLMs and what we should be building instead". URL: https://youtu.be/cEyHsMzbZBs

Table 1: Incomplete answers (Risk 1)

| Prompt-1 -> Key performance indicators show performance achieved and help track the progress of an activity. Can you list all the indicators mentioned in the latest IPCC AR6 reports? ||
|---|---|
| Chat system | Response |
| OpenAI ChatGPT3.5 (OpenAI, model gpt3.5-turbo, 2023)<br><br>Underlying model: GPT3.5-turbo<br><br>Option: None | I apologise for any misunderstanding, but I don't have real-time access to external sources such as the latest IPCC AR6 report, as my training only includes data up until September 2021. Therefore, I can't provide you with a list of indicators mentioned in that specific report. However, I can certainly help you understand and generate examples of common key performance indicators (KPIs) used in various fields if that would be helpful to you. Just let me know what kind of indicators you're interested in! |
| ChatClimate.ai_GPT3.5-turbo_hybrid (ChatClimate, 2023)<br><br>Model: GPT3.5-turbo<br><br>Option Hybrid: Retrieves answers from IPCC reports and in-house GPT knowledge. | There is not enough information to answer the question. |
| ChatClimate.ai_GPT4_hybrid (ChatClimate, 2023)<br><br>Model: GPT4<br><br>https://www.chatclimate.ai/<br><br>Option Hybrid: Retrieves answers from IPCC reports and in-house GPT knowledge. | The specific indicators mentioned in the IPCC AR6 report are not explicitly listed in the information provided. However, it is mentioned that key indicators have been selected across the atmospheric, oceanic, cryospheric and biospheric domains, with land as a cross-cutting component. These indicators are Essential Climate Variables as defined by the Global Climate Observing System (Page:26, Reference: IPCC_AR6_WGI_Chapter02). It's also noted that a wide variety of indicators have been used to assess adaptation and its results, but the specific indicators are not listed (Page:69, Reference: IPCC_AR6_WGII_Chapter17). There is not enough info to answer the question in detail, as the specific indicators are not listed in the information provided. |
| ChatClimate.ai_GPT4_standalone (ChatClimate, 2023)<br><br>Model: GPT4<br><br>Option Standalone: Retrieves answers only from IPCC reports. | The specific indicators mentioned in the IPCC AR6 report include those that capture measures of processes that drive adaptation readiness, including leadership, institutional learning, and intersectoral collaboration. Outcome measures such as the presence of programming known to reduce risks are also mentioned. Indicators related to scaling up of effective interventions and relying on the implementation of science frameworks are considered important. Additionally, measuring impacts attributable to climate change could be addressed with a combination of indicators related to overall health system performance and population vulnerability (Page:74, Reference: IPCC_AR6_WGII_Chapter07). Specific indicators can also be included in the performance metrics and monitoring frameworks for each sector, policy intervention, and budget planning cycle (Page:26, Reference: IPCC_AR6_WGII_Chapter09). However, the report does not provide a comprehensive list of all the indicators used (Page:69, Reference: IPCC_AR6_WGII_Chapter17). |
| IPCC Chat (IPCC Chat, 2023)<br><br>Model: GPT (version unspecified)<br><br>This is an interface to help understand the latest climate change report from the Intergovernmental Panel on Climate Change (IPCC). | Sorry, I don't know how to help with that. The IPCC AR6 report does not provide a comprehensive list of key performance indicators. The report primarily focuses on summarizing the state of knowledge of climate change, its widespread impacts and risks, and climate change mitigation and adaptation, based on the peer-reviewed scientific, technical, and socio-economic literature since the publication of the IPCC's Fifth Assessment Report (AR5) in 2014. The report also identifies opportunities for transformative action which are effective, feasible, just and equitable using concepts of systems transitions and resilient development pathways. |

The chat-based systems did not provide a satisfactory answer to the question (prompt-1). Data sources used in the training of the GPT3.5 Turbo Model do not include IPCC AR6, highlighting the importance of downstream task optimization, such as model fine-tuning based on IPCC AR report data.

The ChatClimate website aims to make climate information accessible to a broader audience. Initially, ChatClimate relied on a Roberta-based pre-trained model publicly available from *hugging-face*. However, the developers of this system now use OpenAI models in replacement of their original pre-trained models.

The ChatClimate.ai_GPT3.5-turbo_hybrid system indicated it did not have enough information to answer the question. When GPT4 models are used, the ChatClimate system provides information on key indicators across atmospheric, oceanic, cryospheric, and biospheric domains. These indicators are essential climate variables defined by the Global Climate Observing System and described on page 26, Chapter 2 of the IPCC AR6 WG I report: The Physical Science Basis (Gulev, 2021). However, the response finished stating that this report "does not hold enough details" to answer the question, although a detailed list of indicators is available on page 27 of the said report.

This example highlights the importance of providing additional contextual information through model fine-tuning or prompt engineering. The Chatclimate system has provided users with two options for answer retrieval: hybrid (which uses IPCC reports and in-house GPT knowledge) and standalone (which only uses IPCC reports). The answers provided differ depending on which option is selected. The standalone option provides information on outcome measures and indicators for measuring impacts. Both options, however, point to the same location in the IPCC AR6 WGII report: https://www.ipcc.ch/report/ar6/wg1/downloads/report/IPCC_AR6_WGI_Chapter02.pdf.

The IPCC Chat system stated that the IPCC AR6 reports do not contain any list of key performance indicators. The IPCC Chat system has not disclosed which version of the GPT model it uses.

## Risk 2: Hallucinations in LLM models for scientific data

The Transformer-based language models have great potential to transform our society. These models can boost generative AI systems that perform text classification, analysis, categorization, closed question-answering (QA) tasks, and abstractive summarization. However, regardless of their potential benefits, assessing the associated risks with these generative AI systems is essential.

In November 2023, the MIT Technology Review (Douglas, 2022) reported on Meta Galactica, an LLM-based system designed to help scientists summarize academic papers and "generate" articles. The scientific community has heavily criticized this system, which shut down its public demo a few days after its launch. Galactica is a large language model trained on more than 40 million examples of scientific data, including academic papers, textbooks, lecture notes, and encyclopedias. The primary issue with Galactica is its inability to distinguish between truth and falsehood. Scientists shared examples of fake scientific articles, real authors associated with unreal papers, and false statements in fabricated Wiki articles.

The analysis of risks associated with LLMs needs to be thorough and provide a comprehensive assessment of the likelihood and impact of risks and the measures that can be taken to mitigate them. For example, LLMs can be deployed as a central or supporting knowledge base of decision-making systems. To ensure the safe and transparent use of LLMs, the European Parliament passed the Artificial Intelligence Act (Parliament, 2023) in June 2023, which requires AI system providers and application developers to undertake risk assessments of foundational language models. Generative AI models must comply with transparency requirements, including disclosing AI-generated content, designing the model to prevent the generation of illegal content, and publishing summaries of copyrighted data used for training.

One issue with LLM-based systems is hallucination or delusion in artificial intelligence. Numerous examples of artificial hallucination have been reported, which demonstrate that LLM-based systems are prone to producing unintended text when prompted with specific factual questions. (Ziwei Ji, 2023) recently published a literature review on hallucination in LLMs. This study defines hallucination in natural language processing as "the generated content that is nonsensical or unfaithful to the provided source content." Hallucinated text appears fluent and natural despite being nonsensical and unfaithful to the context provided. It is often challenging to specify or verify the existence of such contexts. For example, Meta Galactica produced hallucinated text during the system's short period of opening to the public.

Ziwei Ji et al. (Ziwei Ji, 2023) have categorised LLM hallucinations into two main types:

**Type 1.** Intrinsic hallucinations - This type of hallucination generates output that contradicts the source content. For instance, if an abstractive summarisation task outputs that the target value for an energy performance indicator is below 5500 kWh, but the source document reports a target value below 2700 kWh for the same indicator, this is an example of intrinsic hallucination.

**Type 2.** Extrinsic hallucinations - This type of hallucination generates output that cannot be verified from the source content (i.e., the output can neither be supported nor contradicted by the source). For example, the system may generate text not mentioned in the source or fabricate sources of unrelated content to support the generated statements.

As a simple example, suppose we prompt an LLM system with climate-related questions that it answers confidently and even provides source references for the content. However, upon closer examination, we find that the response statements are often wrong, and source references do not exist. These are hallucination errors.

To further demonstrate this, we have extracted an example from a recent interview published on the Climate Foresight website on August 24, 2023[3]. In this interview, the researcher "interviewed" the ChatGPT and asked a series of questions on global climate change. We have taken and adapted two questions from the original interaction to illustrate Type 2 extrinsic hallucinations. Table 2 shows the interaction using the OpenAI chat interface and model gpt3.5-turbo on 05/09/2023.

**Table 2:** Interaction between the user and OpenAI ChatGPT (model gpt3.5-turbo)

---

[3] https://www.climateforesight.eu/articles/how-good-is-chatgpt-at-climate-science/

**Prompt-2 ->** *Let's test this approach by looking at a climate-related question together and then checking the sources that you use to answer my question. Could you tell me what the environmental impact of AI generated language models is and provide scientific references in the text and links to the sources mentioned?*

ChatGPT3.5-turbo-> Certainly! The environmental impact of AI-generated language models is a topic of concern due to the significant computational resources required to train and run these models. Let's explore this issue, and I'll provide you with scientific references to support the information.

Environmental Impact of AI-Language Models: AI-generated language models, like GPT-3 and its successors, have notable environmental impacts primarily in two areas:

Energy Consumption: Training large language models requires substantial computational power, including high-performance GPUs and TPUs. These models undergo training processes that can last for weeks or even months, consuming vast amounts of electricity. The energy demand associated with training deep learning models contributes to greenhouse gas emissions.

Carbon Emissions: The electricity used for training and running AI models is often generated from fossil fuels, which release carbon dioxide ($CO_2$) into the atmosphere. As a result, AI language models can indirectly contribute to carbon emissions and global warming.

Scientific References: Here are some scientific references to support the information above:

"The Carbon Footprint of AI: A Major Problem" - A study by researchers at the University of Massachusetts, Amherst estimates the carbon emissions associated with training large language models. Reference: Link to Study: https://arxiv.org/abs/2010.09718

"Energy and Policy Considerations for Deep Learning in NLP" - This research paper discusses the energy consumption and environmental impact of training large NLP models. Reference: Link to Study: https://arxiv.org/abs/1906.02243

"AI and Carbon Emissions: The Impact of Machine Learning Workloads on Climate Change" - An article published in the journal "Frontiers in Energy Research" explores the relationship between AI workloads and carbon emissions. Reference: Link to Article: https://www.frontiersin.org/articles/10.3389/fenrg.2019.00141/full

"Green AI" - A concept and research area focused on developing environmentally friendly AI algorithms and practices to reduce the carbon footprint of AI systems. Reference: Link to Article: https://www.nature.com/articles/s41586-020-03028-0

These scientific references should provide you with a comprehensive understanding of the environmental impact of AI-generated language models and related research and discussions in the field.

**Prompt-3 ->** *The references you have provided for your answer are inaccurate and therefore cannot be verified. This is a problem that I noticed when asking you to provide sources for the information you give on a multitude of different issues. Do you think this is a problem that is endemic to the way you work and what do you think the consequences could be for the spread of misinformation about issues such as climate change?*

ChatGPT3.5-turbo-> I apologize for any inaccuracies in the references provided in my previous response. As an AI language model, I do not have direct access to external sources or the internet to retrieve real-time information or verify specific sources. Instead, I generate responses based on the knowledge I have been trained on, which is up to date as of September 2021.

The issue of providing accurate and verifiable sources is indeed a challenge when using AI language models, and it can have consequences for the spread of misinformation on various topics, including critical issues like climate change. Some potential consequences include:

Misleading Information: Inaccurate or unverifiable references can disseminate misleading information, making it difficult for users to distinguish between reliable and unreliable sources.

Loss of Trust: If users cannot verify the information provided by AI models, it may erode trust in the technology and its ability to provide accurate information.

Confirmation Bias: Users might be more likely to believe information that aligns with their existing beliefs or biases, potentially reinforcing misinformation.

To mitigate these challenges, it's crucial for users to exercise critical thinking and fact-check information provided by AI models. Additionally, developers and researchers are working on improving the accuracy and reliability of AI-generated responses, including better source attribution and information verification.

Ultimately, while AI language models like me can provide valuable information and assistance, it is essential for users to verify and cross-reference information from trusted and up-to-date sources, especially on critical topics such as climate change, to ensure the accuracy of the information they receive.

The ChatGPT system provides satisfactory answers to prompt 2 and prompt 3. The output focuses on energy consumption associated with GPUs and TPUs and is more general regarding carbon emissions. However, the output does not provide factual data and fails to offer figures for the energy used. Table 3 provides an analysis of the answers received.

**Table 3:** Analysis of the answers received for the two prompts.

| ChatGPT scientific reference (original) | Analysis (authors) | Type of hallucination |
|---|---|---|
| "The Carbon Footprint of AI: A Major Problem" - A study by researchers at the University of Massachusetts, Amherst estimates the carbon emissions associated with training large language models. Reference: Link to Study: https://arxiv.org/abs/2010.09718 | The title of the linked paper is actually "Principle of Multiple Point Criticality in Multi-Scalar Dark Matter Models". This paper does not relate to carbon emissions associated with LLMs. Instead, it deals with a different topic, specifically dark matter models. Also, the authors | *Type 2 - extrinsic hallucination.* Wrong, fabricated reference. |
| | are not affiliated with the University of Massachusetts. The system has mistakenly referred to this paper as relevant to carbon emissions associated with LLMs. | |
| "Energy and Policy Considerations for Deep Learning in NLP" - This research paper discusses the energy consumption and environmental impact of training large NLP models. Reference: Link to Study: https://arxiv.org/abs/1906.02243 | The reference provided is accurate. It refers to a paper published at the 57th Annual Meeting of the Association for Computational Linguistics (ACL) in Florence, Italy, in July 2019. The paper estimates the financial and environmental costs of training various neural network models for natural language processing (NLP). | _No hallucination._ |
| "AI and Carbon Emissions: The Impact of Machine Learning Workloads on Climate Change" - An article published in the journal "Frontiers in Energy Research" explores the relationship between AI workloads and carbon emissions. Reference: Link to Article: https://www.frontiersin.org/articles/10.3389/fenrg.2019.00141/full | Linked article: "Triptycene-Based and Amine-Linked Nanoporous Networks for Efficient CO2 Capture and Separation" by Alam A. et al., which was published in Frontiers in Energy Research on December 2019, reports on the synthesis of triptycene-based porous organic polymers (TBPALs) with amine linkers for CO2 capture and separation. It is important to note that this information is unrelated to any previous answers provided, and the paper does not mention artificial intelligence or workloads. | Severe _Type 2 – extrinsic hallucination._ The generated output cannot be verified using the cited article due to the unrelated topics. |
| "Green AI" - A concept and research area focused on developing environmentally friendly AI algorithms and practices to reduce the carbon footprint of AI systems. Reference: Link to Article: https://www.nature.com/articles/s41586-020-03028-0 | Page not found on www.nature.com. To explore the issue, we searched for "Green AI" on www.nature.com which yielded a handful of references. Among these, only one paper titled "The Carbon Impact of Artificial Intelligence" by Dhar, P. (2020) was relevant. https://doi.org/10.1038/s42256-020-0219-9 | _Type 2 - extrinsic hallucination._ It seems that the reference provided does not exist. However, one could try searching the publisher's website or even the web using the keyword "Green AI" to find any relevant connections. The key takeaway here is that the source content provided has no relationship to the answers, which indicates an intrinsic fault in the system. |

The language model works well in extracting information from knowledge bases during the training and optimization. However, the model output is less effective when prompted to provide sources and scientific references.

The references do not follow open science principles, and their metadata does not match most references but one article. Transparency and data-sharing principles are not thoroughly followed, and one reference cannot even be found through the URL provided (type 2 – extrinsic hallucination). The quality of the output is questionable, and it does not provide factual data from the references (type 1 – intrinsic hallucination).

More precise, coherent, and trustworthy answers require accurate references and quantification. The model hallucinates answers and mistakes when it comes to sources. Only one of the provided references is related to the ChatGPT output, whereas the other three are nonexistent or cover topics unrelated to prompt-1 and prompt-2: energy use and environmental impacts of AI-generated language models.

Further improvements in the models, such as fine-tuning and prompt engineering, require extra contextual data from the user. A user-driven system architecture design could automate knowledge extraction from climate-related texts with an entire user-validation loop and ensure compliance with open science and FAIR principles, which become guidelines and tools within a scientific language context.

## Risk 3: Misinformation

In the context of climate-related data sharing, we categorise data sources into two types. Firstly, trustworthy data is based on good faith and accurate facts and evidence, including reputable scientific publications and high-quality international reports.

Misinformation data, on the other hand, refers to false or inaccurate data that misrepresents the facts. Although misinformation data does not necessarily carry a specific harmful intent, it can still lead people to be misinformed about climate change facts, which can result in harmful decisions for themselves and society at large.

If a trusted entity, such as the government or scientific community, provided today a publicly accessible LLM specialised in one domain, users could be misinformed by currently error-prone LLM technologies. This can be risky if users act upon misinformation, affecting trust in authorities and raising questions about responsibility and liability.

Unfortunately, the web contains intentional and non-intentional errors and outdated information about climate change causes, impacts, adaptation, and mitigation. This misinformation is also incorporated into training data. In principle, generative AI models can be adjusted to suppress this type of error, but the probability of generated outputs from the currently available LLMs still showing bias and errors is high. This is due to a lack of a deeper understanding of this issue among model developers and the research community.

## 3. Conclusions

The potential benefits of using large language models (LLMs) for the climate domain are clear, but it is important to exercise caution. AI-generated texts may not always be accurate, and incomplete answers, hallucinations, and misinformation can be problematic, as discussed in this report.

LLMs should only be considered to enhance the productivity of experts, knowledge curators, and other users who understand the risks and are incentivised to evaluate and improve the AI-generated outputs.

Collaboration between users and LLMs is key to maximizing the potential of these systems and improving their knowledge base. Users can validate the quality of model outputs by providing feedback during inference time (using tailored prompts) or during model training via reinforcement learning from human feedback RLHF.

Also, sharing annotated data among users, such as prompt text formulations, can help bring users back into the loop and extend and improve their knowledge of these models.

## Acknowledgements

Work leading to this paper has been performed in the scope of the EU-funded project MAIA, under the grant agreement ID: 101056935.